\documentclass[conference]{IEEEtran}
\usepackage{ifpdf}

\ifCLASSINFOpdf
   \usepackage[pdftex]{graphicx}
   \graphicspath{{./pdf/}{../jpeg/}{../eps/}}
  \DeclareGraphicsExtensions{.pdf,.jpeg,.png,.eps}
  
\else
   \usepackage[dvips]{graphicx}
   \graphicspath{{./eps/}}
   \DeclareGraphicsExtensions{.eps}
\fi

\usepackage[cmex10]{amsmath}

\usepackage[vlined, ruled, boxed]{algorithm2e}

\usepackage[caption=false,font=footnotesize]{subfig}
\usepackage{url}

\usepackage{booktabs}

\usepackage{multirow}
\usepackage{colortbl}
\usepackage{amssymb}

\begin{document}
\title{The message does not matter: the influence of the network on information diffusion}



\author{\IEEEauthorblockN{Fabr\'{i}cio Olivetti de Franca, \textit{Member, IEEE}}
\IEEEauthorblockA{Universidade Federal do ABC (UFABC) \\ Center of Mathematics, Computing and Cognition (CMCC)\\
R. Santa Ad\'{e}lia 166, CEP 09210-170, Santo Andr\'{e}, Brazil\\
Email: folivetti@ufabc.edu.br}
\and
\IEEEauthorblockN{Denise Hideko Goya}
\IEEEauthorblockA{Universidade Federal do ABC (UFABC) \\ Center of Mathematics, Computing and Cognition (CMCC)}
\and
\IEEEauthorblockN{Claudio Luis de Camargo Penteado}
\IEEEauthorblockA{Universidade Federal do ABC (UFABC) \\ Center of Engineering and Social Sciences (CECS)}

}



\maketitle

\begin{abstract}

How an information spreads throughout a social network is a valuable knowledge sought by many groups such as marketing enterprises and political parties. If they can somehow predict the impact of a given message or manipulate it in order to amplify how long it will spread, it would give them a huge advantage over their competitors. Intuitively, it is expected that two factors contribute to make an information becoming viral: how influential the person who spreads is inside its network and the content of the message. The former should have a more important role, since people will not just blindly share any content, or will they? In this work it is found that the degree of a node alone is capable of accurately predicting how many followers of the seed user will spread the information through a simple linear regression. The analysis was performed with five different messages from Twitter network that was shared with different degrees along the users. The results show evidences that no matter the content, the number of affected neighbors is predictable. The role of the content of the messages of a user is likely to influence the network formation and the path the message will follow through the network.

\end{abstract}

\begin{IEEEkeywords}
complex networks, information diffusion, diffusion model.
\end{IEEEkeywords}

%
\IEEEpeerreviewmaketitle

\section{Introduction}

The influence of a Social Network over an individual behavior can help us understand how a given information is spread. The meaning of information in this context is anything that can be passed from one individual to another, such as, viral diseases~\cite{dodds2005generalized}, political opinion~\cite{pang2008opinion}, innovation adoption~\cite{rogers2005complex}, news and ideas~\cite{bakshy2012role}.

By understanding the mechanisms of information diffusion, we can act in order to prevent the spread of contagious diseases, accelerate the adoption of healthier behavior, and even disseminate an opinion throughout the network. So, an accurate model that can predict such behavior is sought by many researchers from different fields.

A diffusion model~\cite{valente1995network} for a network tries to find the probability that a node spreads a given information to a neighbor node and the probability that a given node absorbs such information from one of its neighbors. For example, in a disease diffusion model, the first probability is related to how infectious a disease is, and the second probability relates to how strong an individual immune system is.

In most cases, these probabilities are assumed to be the same for every node throughout the network, thus reducing the task to estimate only two values per model. This simplification is necessary for many models since the data acquisition is unfeasible. With these values, it is possible to estimate the expected coverage of spread of a given information through a Monte Carlo method.

But, in reality, each individual has its own probabilities on a diffusion model. These probabilities depends on its network position and the content of the message being spread. In a social context of opinion adoption, a person who has contact with many other persons is more likely to spread its opinion through the network, likewise, depending on the opinion being transmitted, a person may be more or less susceptible to retransmit it.

With the appearance of Online Social Networks, it became possible to acquire the data needed for a more thorough investigation of the diffusion model on social networks. One of such networks, named Twitter~\footnote{\url{https://www.twitter.com}}, is a network specifically created to spread opinions and informations~\cite{kwak2010twitter}. This is a directed network where each user may have two possible relationships with another user: follower, when user A listens to what user B says; and friendship, when user A is listened by user B. Every user can publish short messages (140 characters), called tweets, that will show up in its followers pages. The followers are capable of sharing any tweet by means of a retweet. A retweet will show a message created by someone else into the followers pages of the user who retweeted.

The data generated by these interactions is publicly available through the use of developers API with some limitations.

So, by knowing the social network where the message is being spread and the content of such message it may be possible to predict how far the information will reach and how many individual nodes will adopt such information. In this work we will propose a simple model to estimate the rate that any Twitter user spreads any given information based solely on the characteristics of his ego-centric network. Surprisingly, the number of connections alone is capable of accurately predict the average number of followers that will share a given user message.

The remainder of this paper is organized as follows: in Section~\ref{sec:diffusion} we will explain how to generate the retweet diffusion network given the API limitations. Section~\ref{sec:contagion} elaborates the contagion model proposed and the methods to find the spreading rate. Section~\ref{sec:results} follows through a series of experiments in order to determine the best parameters to use for the proposed model. Finally, Section~\ref{sec:conclusion} concludes this papers with some insights for future work.

\section{Retweet Diffusion Network}
\label{sec:diffusion}

The Retweet Diffusion Network (RDN) is a directed network modeled after the dynamics of the sharing mechanism of a tweet message. In this network, the first node is the seed user that originally created the message. Subsequent nodes are created representing the users that shared such message. Each edge from this network represents that a given user $A$ shared the message from another user $B$, not necessarily the seed user.

In Twitter public API~\footnote{\url{https://dev.twitter.com/}}, every tweet is associated with a given user. Whenever someone retweet a message, a new tweet is created with the same content associated with the user who retweeted it. This new message will have a flag indicating this is a retweet (RT) and, additionally, who originally tweeted such message.

This information alone is insufficient to build the correct RDN, since every user who retweeted a given message will be connected with the original user. So, the generation of the RDN must be estimated through the relationships between the users who retweeted the original message.

The first step is to collected the set of users who retweeted the original message, these will be the nodes of the RDN. After retrieving nodes, it is necessary to retrieve the friends list of each node in order to verify from whom they retweeted the message. If a node has only one friend inside the RDN, an edge connecting these nodes will be created. Otherwise, it must be decided from who of his friends the message was retweeted.

In order to estimate the most likely friend from which a given user retweeted, we must hypothesize how the average user reads its Twitter messages. Whenever you open the Twitter website or any of its Smartphone Apps, a list of tweets is shown sorted by the most recent to the least recent. Sometimes, in between these tweets, Twitter shows a trending tweet from the past that might interested the user.

So, the user may have the following behaviors: read a few tweets starting at the most recent and retweet those that it found interesting, read a few tweets from the least recent~\footnote{in this case the user first scroll down the webpage and then slowly scrolls up to the top, while reading the tweets.}, retweet one of the trending tweets that appear among the most recents.

Following these hypothesis, the simplest rule to decide from which user a message was retweeted is to create an edge with the friend who retweeted the message last, thus appearing at the top of the user webpage interface. This will be named RULE 1 for further reference.

Another possible rule is to create an edge to the user with most followers, that retweeted the message within a time frame. This rule makes two suppositions: i) Twitter will give preference to show tweets from the user's more popular friends and, ii) Twitter will not show messages older than a certain amount of time. This rule will be known as RULE 2 from now on.

Finally, the last rule is to create an edge to the user with the least number of followers, that retweeted the message within a time frame. This is the opposite of the previous rule, stating that Twitter will prefer to show the least popular friend. This rule will be named RULE 3.

So, two other hypothesis are created to test such cases. The first hypothesis is that the user will retweet from the user with the most followers given that it was tweeted within a given timeframe (RULE 2). Finally, the last hypothesis is that the user will retweet from its friend with the least followers within a timeframe (RULE 3).

The algorithm used to build the RDN, given the set of users who participated together with the list of their friends who also belong to the RDN and the time of retweet is depicted in Alg.~\ref{alg:RDN}. 

\begin{algorithm}
\caption{RDN building algorithm}
\DontPrintSemicolon
  \SetKwInOut{Input}{input}\SetKwInOut{Output}{output}
\Input{list of nodes $nodes$, associative array of friends for each node $friends$, associative array of time of the retweet for each node $timeRT$, empty graph $G$.}
\Output{Retweet Dynamics Network $G$}
\BlankLine
RDN(nodes,friends,timeRT,G)
$sortedUsers \leftarrow Sort(nodes).by(timeRT)$\;
\For{$node \in sortedUsers$}{
  \If{$node \notin G$}{
    $friend \leftarrow Rule(friends[node], timeRT, timeRT[node])$\;
    $G.add\_edge(node,friend)$\;
    $RDN(nodes,friends,timeRT,G)$\;
  }
}
\label{alg:RDN}
\end{algorithm}

The algorithm is a simple recursive depth-first search that starts with an empty graph $G$ and, starting from the first tweet, apply one of the aforementioned rules to choose which friend will connected to the current user. The three possible rules are given in Algs.~\ref{alg:R1},~\ref{alg:R2},~\ref{alg:R3}.

\begin{algorithm}
\caption{Rule 1 of retweet dynamics}
\DontPrintSemicolon
  \SetKwInOut{Input}{input}\SetKwInOut{Output}{output}
\Input{list of possible connections $friends$, associative array of time of the retweet for each node $timeRT$, time of tweeting for current node $timeUser$.}
\Output{Chosen friend $friend$}
\BlankLine
$friend \leftarrow Mininum(friends).by(timeRT-timeUser)$\;
\label{alg:R1}
\end{algorithm}

\begin{algorithm}
\caption{Rule 2 of retweet dynamics}
\DontPrintSemicolon
  \SetKwInOut{Input}{input}\SetKwInOut{Output}{output}
\Input{list of possible connections $friends$, associative array of time of the retweet for each node $timeRT$, time of tweeting for current node $timeUser$.}
\Output{Chosen friend $friend$}
\BlankLine
$Filtered \leftarrow Filter(SortedFriends).by(|timeRT - timeUser| \leq thr)$\;
$friend \leftarrow Max(Filtered).by(NumberOfFriends)$\;
\label{alg:R2}
\end{algorithm}

\begin{algorithm}
\caption{Rule 3 of retweet dynamics}
\DontPrintSemicolon
  \SetKwInOut{Input}{input}\SetKwInOut{Output}{output}
\Input{list of possible connections $friends$, associative array of time of the retweet for each node $timeRT$, time of tweeting for current node $timeUser$.}
\Output{Chosen friend $friend$}
\BlankLine
$Filtered \leftarrow Filter(SortedFriends).by(|timeRT - timeUser| \leq thr)$\;
$friend \leftarrow Min(Filtered).by(NumberOfFriends)$\;
\label{alg:R3}
\end{algorithm}

\begin{figure}[!htb]
\centering
\includegraphics[width=0.45\textwidth]{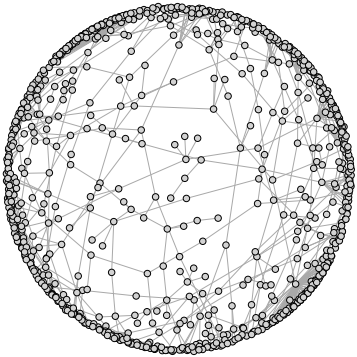}
\caption{An example of a RDN using Rule 1.}
\label{fig:RDN}
\end{figure}

\begin{figure}[!htb]
\includegraphics[width=0.45\textwidth]{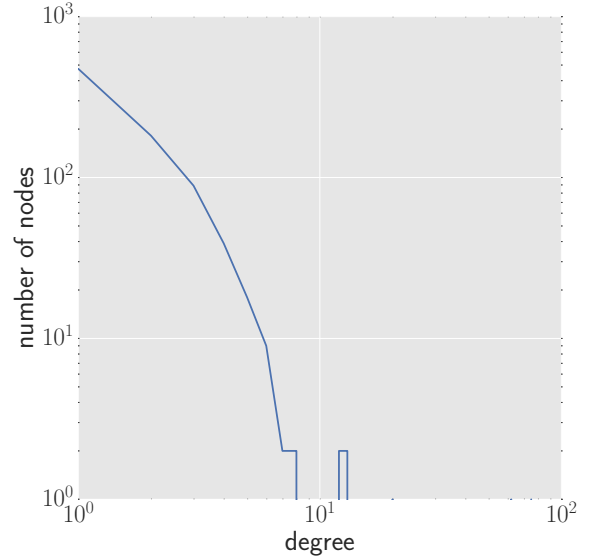}
\caption{Degree distribution in log-log scale.}
\label{fig:degree}
\end{figure}

The first rule is straightforward and just retrieves the friends which tweeted last. The other two rules, first filter those friends who tweeted within a time limit $thr$ and then retrieve the friend with more and less followers, respectively. In the following Sections such rules will be tested for different values of $thr$. An illustration of a RDN when applying the first rule is depicted in~\ref{fig:RDN} with its corresponding degree distribution in~\ref{fig:degree}.

The degree relationship of this network assumes a power law distribution regarding the number of friends that retweeted from a given node. This indicates that the number of potential retweets each user is capable of generating is predictable by a simple linear regression as we will show in the following Sections.

\section{Contagion Model}
\label{sec:contagion}

After the construction of the RDN, the next step is trying to predict how the tweets are shared. In this paper we will use a simple contagion model known as Susceptible-Infected (SI)~\cite{bai2007immunization}. In our context, this model says that initially every node, except for the seed node is on a susceptible state, the seed node is on an infected state, being capable of transmitting the information to its neighbor.

At every time step, the message will spread from every infected node to a fraction $\beta$ of their neighbors:

\begin{equation}
nRT = \beta F(i),
\label{eq:nrt}
\end{equation}

\noindent where $nRT$ is the number of retweets the user $i$ will receive and $F(.)$ is the number of followers a given user has.

In most contagion models the transmission rate is the same for the entire network, in our case each user will have a different transmission rate since different users have a different potential to spread a given message. So, Eq.~\ref{eq:nrt} must be changed to:

\begin{equation}
nRT = \beta(i) F(i),
\label{eq:nrt2}
\end{equation}

The estimation of $\beta(i)$ may depend on different attributes from the user, such as: number of followers, number of friends, number of posts, how long the message has been sent, and the content of the message itself. Many related work~\cite{petrovic2011rt,walther2001impacts,yang2010predicting,lampos2014predicting} focused on the content of the message with the objective of predicting how much a message will spread. But, due to the power-law distribution nature of the nodes degrees, we will try create a linear regression model based on the logarithm of the values of such attributes, except for the textual message:

\begin{equation}
\log\beta(i) = w_1 \dot \log F(i) + w_2 \dot \log Fr(i) + w_3 \dot \log P(i) + w_4 \dot \log T(i),
\end{equation}

\noindent where $F(i)$ is the number of followers of user $i$, $Fr(i)$ is the number of friends of user $i$, $P(i)$ is the number of posts made by user $i$, $T(i)$ represents how much seconds have passed since the original tweet was posted and $w_j$ is the weigh associated to each feature.

In the next section we will make different experiments that will show that just a subset of these parameters are enough to predict the $\beta$ rate for any Twitter user with high accuracy.

\section{Experimental Methods and Results}
\label{sec:results}

This Section will first describe the data acquisition process and the datasets used for the experiments. After that, a series of experiments will be performed in order to find the best model that we can obtain using as few information about the tweet as possible. Finally, the results will be reported together with a brief reporting of related work.

\subsection{Data Acquisition}

In order to verify our assumptions some data were acquired from Twitter through its developer API. This API has several limitations that impedes a thorough experiment. One limitation is the number of requests per $15$ minutes time frame that slows down the data collection process. In pratice this limitation makes unfeasible to collect datasets with more than a few thousands of retweets. Another limitation concerns the age of the tweet being retrieved, the API allows only to search for the tweets being posted during the last week, which limits the dataset collection possibilities.

Because of these limitations, we have collected the information regarding the users, their friends, timestamp, number of friends, number of followers and number posts from $5$ different messages involving $4$ different users. The collection was perform from april 20 to april 29 and corresponds to highly retweeted messages from popular users. Additionally, for one of the users, we acquired a message with a smaller number of retweets in order to test the generality of the proposed model. A brief description of each dataset is given in Table~\ref{tab:dataset}.

\begin{table*}
\centering
\caption{Dataset description.}
\begin{tabular}{c|c|c|c}
\textbf{Name} & \textbf{Tweet} & \textbf{Seed User} & \textbf{\# of RTs.} \\
\hline
RT1 & From "Liam is our batman" & @zaynbaabe & 821 \\
RT2 & Exactly 1 WEEK 'til @ShawnMendes takes over \#VH1Buzz for 5 days! \#ShawnOnBUZZ & @BigMorningBuzz & 2,704 \\
RT3 & Using Direct Messages to take public conversations private is now easier than ever. Learn about today's updates: & @twitter & 921 \\
RT4 & Are you ready for \#XenobladeChroniclesX? Here’s a handy \#XenobladeChronicles3D checklist to prepare you. & @NintendoAmerica & 309 \\
RT5 & Kit and Krysta take on the "Whisper Challenge" in the latest \#NintendoMinute http://Ninten.do/6017fv2Z & @NintendoAmerica & 32 \\
\end{tabular}
\label{tab:dataset}
\end{table*}

\subsection{Linear Regression}

In our model, the value of the retweet rate $\beta$ will be estimated by means of a Least Squares Linear Regression algorithm. The regression was performed using the Scikit-Learn 0.16 library with Python 3.3.5. In order to assess the accuracy of the regression, for every experiment, one dataset was chosen as the training data to be fitted, and the remaining datasets were used as a test set. 

The quality of the obtained solutions was measured by means of the coefficient of determination ($R^2$), the Mean Absolute Error (MAE) and the Mean Squared Error (MSE) of the output. Notice that the errors were calculated regarding the exponential of the output and compared with the measured $\beta$ for each case.


\subsection{Rule of Diffusion}

The first test will verify what hypothesis for creating the RDN gives the most accurate results. For this purpose, a series of experiments were performed in which we create a regression model using different rules as described in Section~\ref{sec:diffusion}.

The tested rules were: Rule 1 (R1), Rule 2 with 15 minutes time frame (R2\_15), with 30 minutes time frame (R2\_30) and with 60 minutes time frame (R2\_60), and Rule 3 with 15, 30 and 60 minutes time frames (R3\_15, R3\_30, R3\_60).

Table~\ref{tab:hypothesis} shows the average result obtained by every rule by training the model with each dataset and testing with the remaining data. In this Table we can see that the rule R2\_15 is the most accurate regarding the average error metrics, but R3\_60 obtained a much better value for $R^2$ while still maintaining low values for MAE and MSE. For this purpose the R3\_60 rule will be used for the next set of experiments.

\begin{table}
\centering
\caption{Results obtained for each hypothesis rule.}
\begin{tabular}{llll}
\toprule
{} &       $\mathbf{R^2}$ & \textbf{MAE} & \textbf{MSE} \\
\midrule
R1    &  $0.846579$ &  $0.001292$ &  $0.000018$ \\
R2\_15 &  $0.832880$ &  $0.001168$ & $0.000008$ \\
R2\_30 &  $0.580089$ &  $0.006751$ &  $0.003070$ \\
R2\_60 &  $0.570107$ &  $0.007283$ &  $0.003457$ \\
R3\_15 &  $0.851444$ &  $0.001328$ &  $0.000016$ \\
R3\_30 &  $0.854654$ &  $0.001305$ &  $0.000015$ \\
R3\_60 &  $0.863555$ &  $0.001277$ &  $0.000014$ \\
\bottomrule
\end{tabular}
\label{tab:hypothesis}
\end{table}

These results indicate that it is more probable that a user retweet from its less connected friend during this timeframe. One possible explanation for these results is that this illustrates a Retweet-Follow behavior. During the retweeting event, the user originally did not follow the most popular user from its current network. But, when the user retweeted the message, it decided to follow the twitter account that originally tweeted the message.

\subsection{The Training Set}

For the next set of experiments, by fixing the use of the R3\_60 rule to generate the RDN, it will be verified which dataset renders the best results when used as the training set. The same set of experiments is performed by varying the dataset used for training while using the remaining as the test data.

As depicted in Table~\ref{tab:train}, the RT1 and RT2 datasets obtained the best values for $R^2$, while RT5 had the best values for the average error metrics, probably caused by the reduced number of samples. Based on these results, RT2 will be chosen as the training set for the remaining experiments. 

\begin{table}
\centering
\caption{Results obtained for each training set.}
\begin{tabular}{llll}
\toprule
{} &       $\mathbf{R^2}$ & \textbf{MAE} & \textbf{MSE} \\
\midrule
RT1 &  0.894002 &  0.001209 &  0.000016 \\
RT2 &  0.891166 &  0.001128 &  0.000011 \\
RT3 &  0.872202 &  0.001362 &  0.000018 \\
RT4 &  0.801393 &  0.001627 &  0.000017 \\
RT5 &  0.859011 &  0.001059 &  0.000007 \\
\bottomrule
\end{tabular}
\label{tab:train}
\end{table}

It might be a surprise that the dataset with larger dataset obtained one of the a highest coefficient of determination. Intuitively, smaller datasets is expected to be easier to fit into a linear regression in comparison with larger datasets, that contain lots of noise. But, in this case, a smaller number of samples might mean that we could not acquire the complete set or that the tweet is still being retweeted along the network, thus increasing the noise.

\subsection{The best feature set}

The final set of experiment will verify which combination of the features set gives the best results. For this purpose all combinations of the features message time (time), number of friends (friends), number of followers (followers) and number of posts (posts) were tested using the RT2 dataset as the training data and all of the datasets as the test set.

Table~\ref{tab:features} does not give us a clear winner on this combination, but we can see that the time and posts features do not seem to be discriminative regarding our data, while the followers attribute alone could achieve a very accurate result. By further inspection, the combination (friends, followers) was capable of obtaining a minimum error with a high coefficient of determination.

As such, this will be the chosen attributes for the final model.

\begin{table}
\centering
\caption{Results obtained for each features subset.}
\begin{tabular}{llll}
\toprule
{} &       $\mathbf{R^2}$ & \textbf{MAE} & \textbf{MSE} \\
\midrule
(time)                         & -0.274353 &  0.004399 &  0.000134 \\
(friends)                      &  0.376648 &  0.003237 &  0.000081 \\
(followers)                    &  0.849871 &  0.000984 &  0.000009 \\
(posts)                        & -0.059239 &  0.004003 &  0.000104 \\
(time,friends)                 &  0.377421 &  0.003240 &  0.000081 \\
(time,followers)               &  0.856084 &  0.000984 &  0.000008 \\
(time,posts)                   & -0.059073 &  0.003999 &  0.000104 \\
(friends,followers)            &  0.880330 &  0.000982 &  0.000009 \\
(friends,posts)                &  0.382251 &  0.003182 &  0.000088 \\
(followers,posts)              &  0.869515 &  0.001275 &  0.000014 \\
(time,friends,followers)       &  0.883551 &  0.001056 &  0.000007 \\
(time,friends,posts)           &  0.381019 &  0.003174 &  0.000088 \\
(time,followers,posts)         &  0.876833 &  0.001164 &  0.000012 \\
(friends,followers,posts)      &  0.885927 &  0.001225 &  0.000013 \\
(time,friends,followers,posts) &  0.891166 &  0.001128 &  0.000011 \\
\bottomrule
\end{tabular}
\label{tab:features}
\end{table}

It is interesting to notice that the time of the retweet is not an important factor to predict the retweet rate of a given user, suggesting that the retweet rate does not attenuate through time. So, in this model, the tweet will only stop being retweeted when it reaches users with a small number of followers.

\begin{figure}[!htb]
\centering
\includegraphics[width=0.48\textwidth]{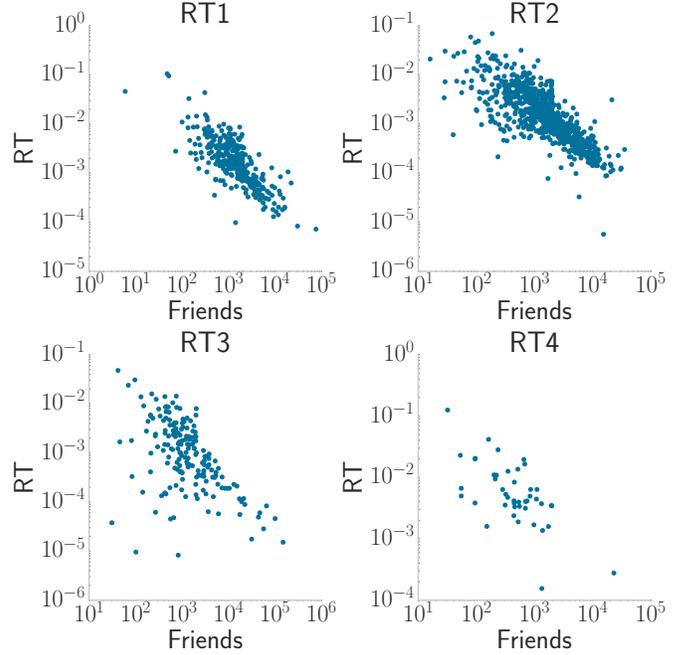}
\caption{Correlation of Number of Friends with the measured $\beta$.}
\label{fig:friends}
\end{figure}

\begin{figure}[!htb]
\centering
\includegraphics[width=0.48\textwidth]{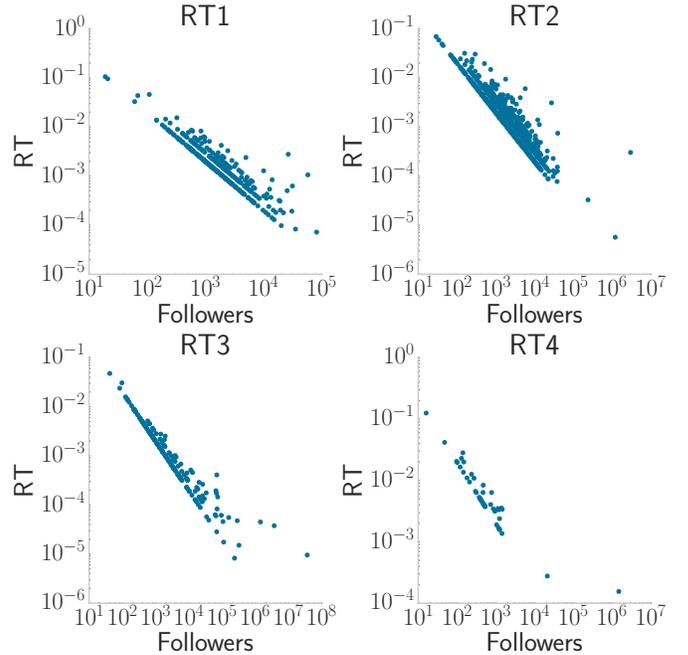}
\caption{Correlation of Number of Followers with the measured $\beta$.}
\label{fig:followers}
\end{figure}

\begin{figure}[!htb]
\centering
\includegraphics[width=0.48\textwidth]{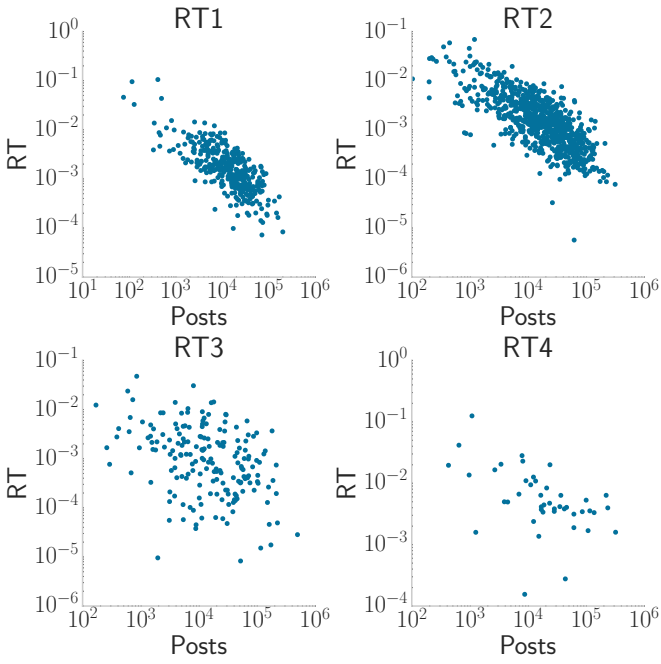}
\caption{Correlation of Number of Posts with the measured $\beta$.}
\label{fig:posts}
\end{figure}

\begin{figure}[!htb]
\centering
\includegraphics[width=0.48\textwidth]{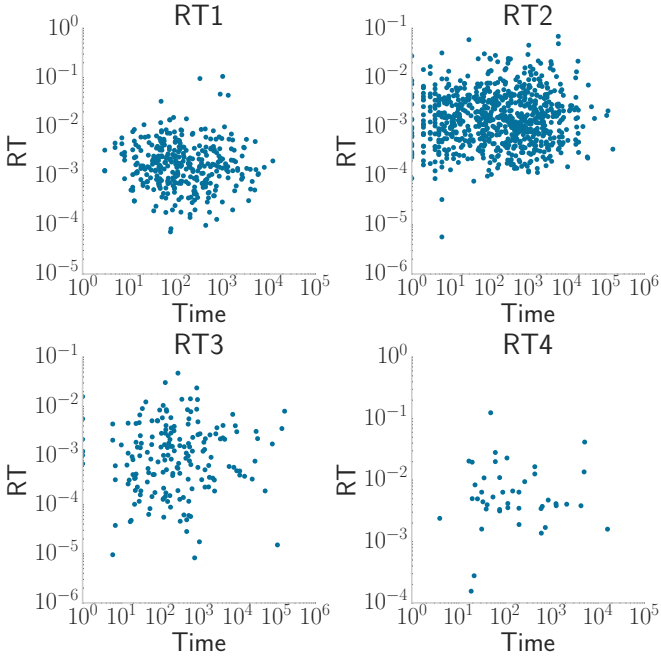}
\caption{Correlation of Time of Retweet with the measured $\beta$.}
\label{fig:time}
\end{figure}

The correlation of each feature with the measured $\beta$ in log-scale is depicted in Figs.~\ref{fig:friends},~\ref{fig:followers},~\ref{fig:posts},~\ref{fig:time}.

From these figures we can confirm that the number of followers has the most well behaved correlation with the measured $\beta$ followed by the number of friends. The number of posts has some correlation on the larger datasets and the time of retweet does not seem to have any correlation at all.

\subsection{The Diffusion Model}

Finally, by fitting the linear regression model with the selected subset of features, the rule to generate the RDN and the largest training set, we get the results in Table~\ref{tab:finalresults}. As we can see the goodness of fit is very high giving us a high confidence on the predictions. This can be verified by the low values of MAE and MSE.

\begin{table}
\centering
\caption{Final results obtained for each dataset.}
\begin{tabular}{llll}
\toprule
{} &       $\mathbf{R^2}$ & \textbf{MAE} & \textbf{MSE} \\
\midrule
RT1 &  0.852532 &  0.000872 &  0.000007 \\
RT2 &  0.869640 &  0.000844 &  0.000004 \\
RT3 &  0.919541 &  0.000452 &  0.000001 \\
RT4 &  0.784082 &  0.002535 &  0.000030 \\
RT5 &  0.975856 &  0.000208 &  0.000000 \\
\bottomrule
\end{tabular}
\label{tab:finalresults}
\end{table}

\begin{figure}[!htb]
\centering
\includegraphics[width=0.48\textwidth]{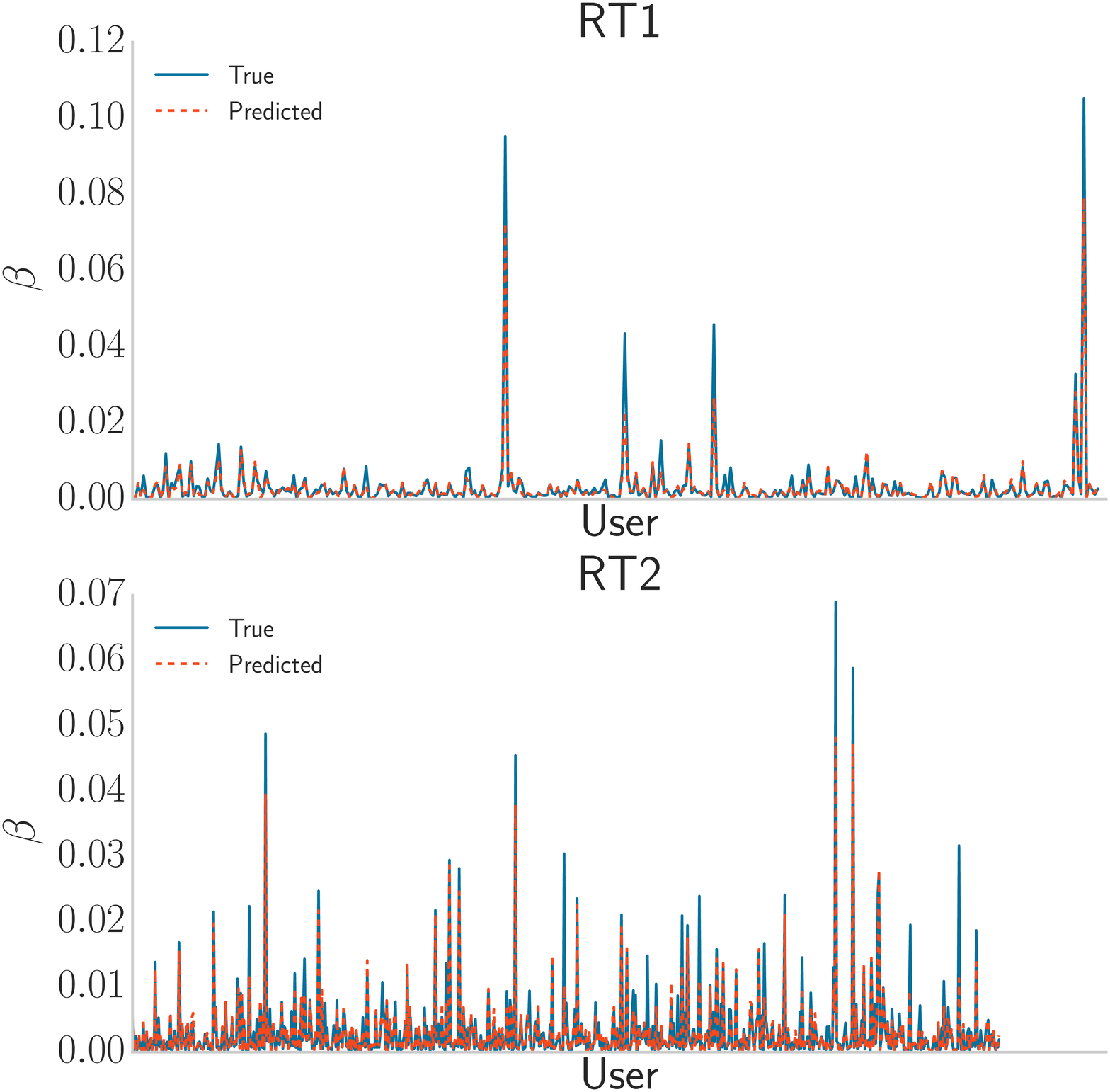}
\caption{True and predicted $\beta$ for RT1 and RT2 datasets.}
\label{fig:fit1}
\end{figure}

\begin{figure}[!htb]
\centering
\includegraphics[width=0.48\textwidth]{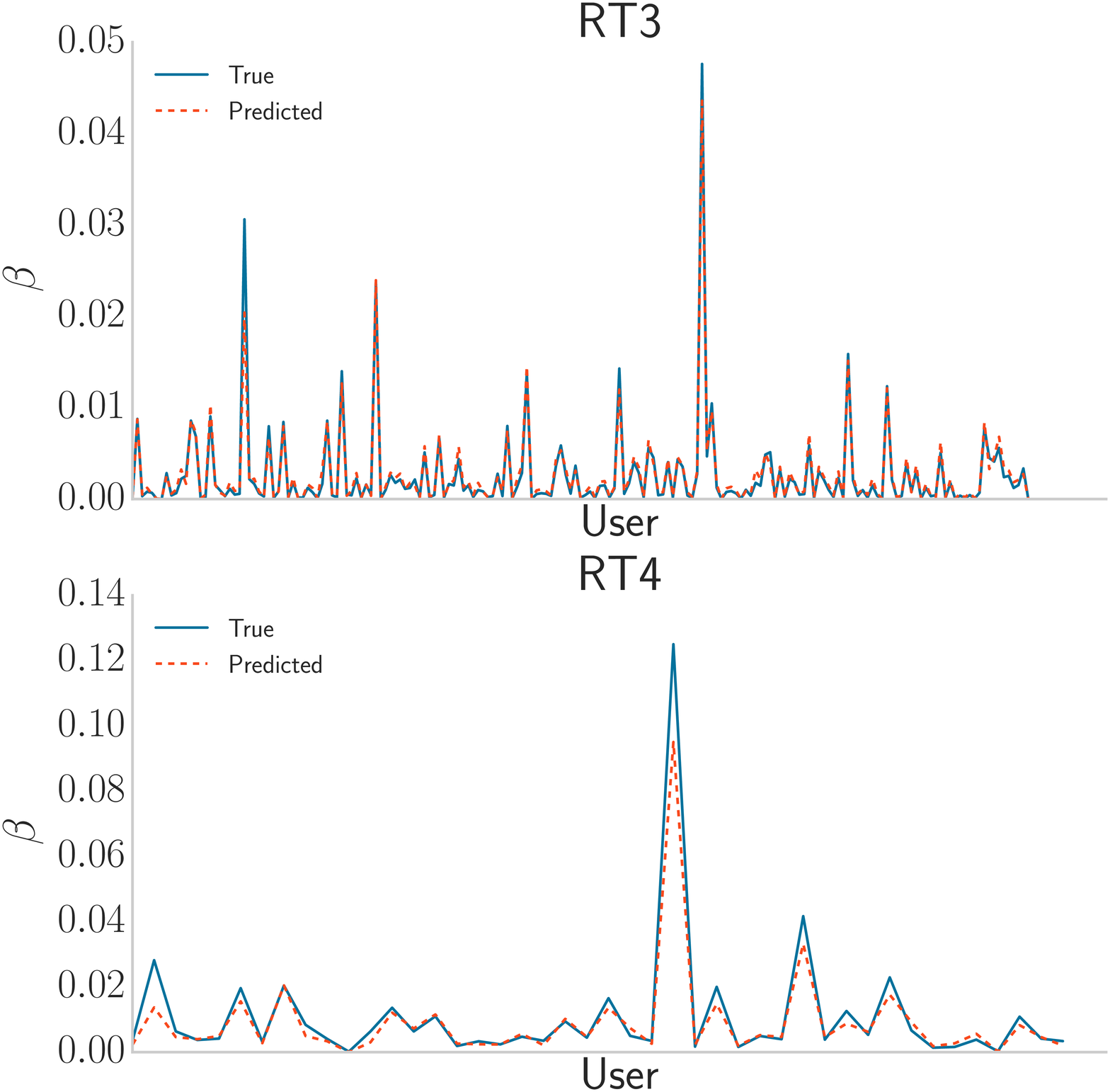}
\caption{True and predicted $\beta$ for RT3 and RT4 datasets.}
\label{fig:fit2}
\end{figure}

A visual depiction of the model accuracy can also be seen in Figs.~\ref{fig:fit1}~and~\ref{fig:fit2}.

As we can see from these plots, the fitted line is very close to the measured values, but underestimates some of the higher impact rates.

The fitted line, after converted back from the log-scale, give us the following equation for $\beta$:

\begin{equation}
\beta = Followers^{-0.77} \times Friends^{-0.12}.
\end{equation}

This equation shows that the number of followers plays an important role to how much retweets a given user will get on average. With less impact, but still important, the number of friends also helps to measure the influence of such user. At this moment, the reader should have already notice that this model, although highly accurate, does not take into account the message being transmitted. So, in the Twitter Network, the messages are being spread regardless of their contents.

One possible explanation can be found in the user behavior to choose its friends. A certain user will choose to follow someone who: is its friend in real life or tweets contents of interest. So the content of the message might play an important role on the network formation.

Additionally, even though we can use this model to determine the fraction of the followers that will retweet your message, we still cannot state which users will most likely do so. This can be important in order to detect the path that a tweet will walk through the RDN and to estimate the depth of such network.

\begin{figure}[!htb]
\centering
\includegraphics[width=0.48\textwidth]{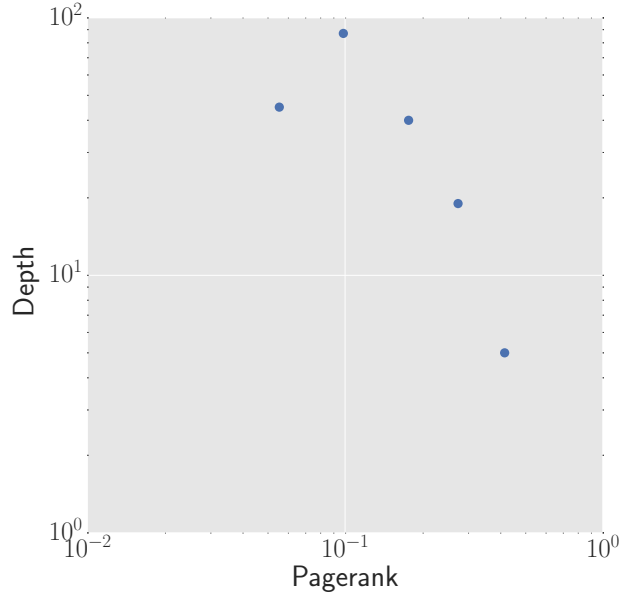}
\caption{Correlation between pagerank of the seed user and the depth of the RDN.}
\label{fig:length}
\end{figure}

The diameter and the average path length for every tree of our proposed model is reported in Tab.~\ref{tab:length}. From this table we can see that the depth of the information diffusion network is closely related to the number of retweets it received. In Fig.~\ref{fig:length} we have plotted the estimated Pagerank of the seed user on the RDN and the depth of the obtained network in log-log scale. This graphic corroborate with our intuition that the pagerank of the seed user is related to the impact of the message diffusion.

\begin{table}
\centering
\caption{Depth and average path length for each network. |E| represents the number of edges.}
\begin{tabular}{lrrr}
\toprule
{} &   \textbf{|E|} &  \textbf{Depth} &  \textbf{avg. path length} \\
\midrule
RT1 &   $821$ &     $45$ &         $18.52$ \\
RT2 &  $2,704$ &     $87$ &         $26.23$ \\
RT3 &   $921$ &     $40$ &          $9.21$ \\
RT4 &   $309$ &     $19$ &          $3.79$ \\
RT5 &    $32$ &      $5$ &          $2.26$ \\
\bottomrule
\end{tabular}
\label{tab:length}
\end{table}

\subsection{Related Work}
\label{sec:related}

Although we cannot compare these results directly with different models, we can verify how successful other models were through the coefficient of determination.

The most related work was done in~\cite{yang2010predicting} where the authors tracked the diffusion of a message through the mention of a particular user. The network was built by connecting two users, A and B, if user A has mentioned B on a similar topic that B tweeted about. After that, they ran a regression analysis trying to correlate the number of posts, number of mentions a user make, number of mentions an user receive, and some other related information. With these variables, the authors tried to predict when one user would mention another one, the number of connections a user will have in their netwrok and the diameter of the generated network. For the diameter and the time of mention they obtained a coefficient of determination below $0.1$, while for the number of connections, closely related to our RT rate, obtained an $R^2$ in the range of $0.15$ to $0.41$.

In~\cite{petrovic2011rt} they use a Passive-Agressive algorithm to classify whether a message will be retweeted or not. This algorithm that tries to learn a linear boundary between two classes. For this purpose, they used the following variables: number of friends, number of followers, number of posts, favorites, number of times the user was listed, whether the user is verified by Twitter, and if the user's language was English. Additionally, the added the number of hashtags of the tweet, number of mentions, trending words, length of tweet, novelty, if the tweet was a reply and the bag-of-words of the tweet. They measured the quality of their approach through the $F_1$ score that calculates the harmonic mean between the correct and incorrect classification. Their results cannot be compared directly with our approach, but their numerical results ($F_1 = 46.6$) suggest that they still have a large margin of incorrect classification, while our approach is close to the true value (see Figs.~\ref{fig:fit1}~and\ref{fig:fit2}).

Another related work was performed recently in~\cite{lampos2014predicting} where the authors apply linear and non-linear regression models to predict a user impact score. The user impact was measured as the logarithm of the number of followers multiplied by the number of lists and divided by the number of friends. Then, they tried to predict this impact measure with different regression models using textual and non-textual attributes. The maximum obtained coefficient of determination for the non-linear regression was $0.78$, lower then our average results.

Also, closely related, in~\cite{bakshy2011everyone} the authors created a similar diffusion network by following three different rules: create and edge between the user who retweeted to the friend that tweeted last, create an edge between the user who retweeted and the friend that tweeted first, create an edge from the user who retweeted to every friend that tweeted that message, distributing the weight evenly. They reported the results just for the first rule, since they found that there was no significant difference among them. Then, they fitted their data with a Regression Tree Model using the same attributes described here combined with the average, minimum and maximum local and global influences. Local influence in this paper is the number of friends that retweeted a message, while global influence is the number of users of Twitter that retweeted such message. The measured output in this work was the global influence. They obtained a $R^2$ of $0.98$ regarding the most influent users and $R^2$ of $0.34$ when fitting just the least influent users.

\section{Conclusion}
\label{sec:conclusion}

In this paper we proposed a diffusion model for the retweet behavior of the Twitter network. This model is capable of predicting the rate of the followers of a given user that will retweet its messages.

At first, a Retweet Diffusion Network was created by connecting the users with a relationship of who retweeted from who. Some edges of such network had to be estimated because of the lack of information provided by the Twitter application interface. Three proposals of estimation was evaluated and the most plausible was that a user will retweet a message from his friend with the smaller number of connections.

After generating such network, the impact of each user was measured by the fraction of its followers that retweeted its message. This impact was fitted through a series of different linear regression models by using the user social network information such as: number of followers, number of friends, number of posts and time of the post. Since a power-law fit was observed regarding the degree distribution of the Diffusion Network, the regression model was performed on the logarithm of the parameters value and the desired output value.

Some experiments were performed on $5$ different networks with a varying number of nodes and edges starting from different seed users. The results showed that the combination of the number of followers and number of friends is capable of predicting the expected fraction of followers that will retweet a message from any given user.

These results indicate that the content of the message has no influence on the immediate impact. The content must have a central role during the network formation (who follows who) and the path the message will take (who from the fraction of users will spread the message).

The next steps of this research will try to evaluate the decision process that drives any given user to retweet a message, hopefully finding a model to predict the path a message will go through. If successful, it will be possible to predict the global impact of a tweet message and how to build a network to influence the desired audience.

\section*{Acknowledgment}

This research was funded by FAPESP process number 2014/06331-1.

\IEEEtriggeratref{18}

\bibliographystyle{IEEEtran}
\bibliography{TheMessageDoesNotMatter}

\begin{thebibliography}{10}
\providecommand{\url}[1]{#1}
\csname url@samestyle\endcsname
\providecommand{\newblock}{\relax}
\providecommand{\bibinfo}[2]{#2}
\providecommand{\BIBentrySTDinterwordspacing}{\spaceskip=0pt\relax}
\providecommand{\BIBentryALTinterwordstretchfactor}{4}
\providecommand{\BIBentryALTinterwordspacing}{\spaceskip=\fontdimen2\font plus
\BIBentryALTinterwordstretchfactor\fontdimen3\font minus
  \fontdimen4\font\relax}
\providecommand{\BIBforeignlanguage}[2]{{%
\expandafter\ifx\csname l@#1\endcsname\relax
\typeout{** WARNING: IEEEtran.bst: No hyphenation pattern has been}%
\typeout{** loaded for the language `#1'. Using the pattern for}%
\typeout{** the default language instead.}%
\else
\language=\csname l@#1\endcsname
\fi
#2}}
\providecommand{\BIBdecl}{\relax}
\BIBdecl

\bibitem{dodds2005generalized}
P.~S. Dodds and D.~J. Watts, ``A generalized model of social and biological
  contagion,'' \emph{Journal of Theoretical Biology}, vol. 232, no.~4, pp.
  587--604, 2005.

\bibitem{pang2008opinion}
B.~Pang and L.~Lee, ``Opinion mining and sentiment analysis,''
  \emph{Foundations and trends in information retrieval}, vol.~2, no. 1-2, pp.
  1--135, 2008.

\bibitem{rogers2005complex}
E.~M. Rogers, U.~E. Medina, M.~A. Rivera, and C.~J. Wiley, ``Complex adaptive
  systems and the diffusion of innovations,'' \emph{The Innovation Journal: The
  Public Sector Innovation Journal}, vol.~10, no.~3, pp. 1--26, 2005.

\bibitem{bakshy2012role}
E.~Bakshy, I.~Rosenn, C.~Marlow, and L.~Adamic, ``The role of social networks
  in information diffusion,'' in \emph{Proceedings of the 21st international
  conference on World Wide Web}.\hskip 1em plus 0.5em minus 0.4em\relax ACM,
  2012, pp. 519--528.

\bibitem{valente1995network}
T.~W. Valente, \emph{Network models of the diffusion of innovations}.\hskip 1em
  plus 0.5em minus 0.4em\relax Hampton Press Cresskill, NJ, 1995, vol.~2,
  no.~2.

\bibitem{kwak2010twitter}
H.~Kwak, C.~Lee, H.~Park, and S.~Moon, ``What is twitter, a social network or a
  news media?'' in \emph{Proceedings of the 19th international conference on
  World wide web}.\hskip 1em plus 0.5em minus 0.4em\relax ACM, 2010, pp.
  591--600.

\bibitem{bai2007immunization}
W.-J. Bai, T.~Zhou, and B.-H. Wang, ``Immunization of susceptible--infected
  model on scale-free networks,'' \emph{Physica A: Statistical Mechanics and
  its Applications}, vol. 384, no.~2, pp. 656--662, 2007.

\bibitem{petrovic2011rt}
S.~Petrovic, M.~Osborne, and V.~Lavrenko, ``Rt to win! predicting message
  propagation in twitter.'' in \emph{ICWSM}, 2011.

\bibitem{walther2001impacts}
J.~B. Walther and K.~P. D’Addario, ``The impacts of emoticons on message
  interpretation in computer-mediated communication,'' \emph{Social science
  computer review}, vol.~19, no.~3, pp. 324--347, 2001.

\bibitem{yang2010predicting}
J.~Yang and S.~Counts, ``Predicting the speed, scale, and range of information
  diffusion in twitter.'' \emph{ICWSM}, vol.~10, pp. 355--358, 2010.

\bibitem{lampos2014predicting}
V.~Lampos, N.~Aletras, D.~Preo{\c{t}}iuc-Pietro, and T.~Cohn, ``Predicting and
  characterising user impact on twitter.''\hskip 1em plus 0.5em minus
  0.4em\relax Association for Computational Linguistics, 2014.

\bibitem{bakshy2011everyone}
E.~Bakshy, J.~M. Hofman, W.~A. Mason, and D.~J. Watts, ``Everyone's an
  influencer: quantifying influence on twitter,'' in \emph{Proceedings of the
  fourth ACM international conference on Web search and data mining}.\hskip 1em
  plus 0.5em minus 0.4em\relax ACM, 2011, pp. 65--74.

\end{thebibliography}

\end{document}